\documentclass[aps,pra,twocolumn,showpacs,groupedaddress,amsmath,amssymb]{revtex4}

\usepackage{graphicx}
\graphicspath{{b:/officestuff/papers/nonlin_LRC/figures/}}

\usepackage{color}
\usepackage{comment}
\usepackage{bm}
\usepackage{amsmath}
\usepackage{latexsym}

\begin{document}

\title{Wavepacket Self-imaging and Giant Recombinations via Stable Bloch-Zener Oscillations in Photonic Lattices with Local ${\cal PT}$-Symmetry}
\author{N. Bender,  H. Li, F. M. Ellis, T. Kottos}
\affiliation{Department of Physics, Wesleyan University, Middletown, CT-06459, USA}
\date{\today}

\begin{abstract}

We propose a family of {\it local} $\cal{PT}$-symmetric photonic lattices with transverse index gradient $\omega$, where the
emergence of {\it stable} Bloch-Zener oscillations are controlled by the degree of non-Hermiticity $\gamma$ of the lattice. In
the exact $\cal{PT}$-symmetric phase we identify a condition between $\omega$ and $\gamma$ for which a wavepacket self
-imaging together with a cascade of splittings and giant recombinations occurs at various propagation distances. The giant 
wavepacket recombination is further enhanced by introducing local impurities.
\end{abstract}
\maketitle

{\it Introduction -} Non-Hermitian wave physics and specifically its parity-time (${\cal PT}$) symmetric ramifications \cite{BB98}, has attracted a lot of attention in recent years. The main observation was that a non-Hermitian Hamiltonian ${\cal H}$ that commutes with the joint ${\cal PT}$-symmetric operator may possess an entirely real spectrum. Specifically it was shown that below a critical value $\gamma_{\cal PT}$, of the parameter $\gamma$ controlling the non-Hermiticity of ${\cal H}$, the spectrum is real and the eigenfunctions of ${\cal H}$ are eigenfunctions of the ${\cal PT}$-symmetric operator. In the opposite limit the spectrum becomes partially or completely complex while the eigenfunctions cease to be eigenfunctions of the ${\cal PT}$ operator. The first
domain was coined the exact ${\cal PT}$-symmetric phase while the latter was coined the broken ${\cal PT}$-symmetric phase. The transition point $\gamma=\gamma_{\cal PT}$ is known as an exceptional point (EP) singularity where both the eigenfunctions and eigenvalues coalesce.

The impact of these ideas is well documented in various physical settings ranging from matter waves \cite{CW12,GKN08} and magnonics \cite{LKS14} to optics \cite{MGCM08,RMGCSK10,FXFLOACS13,GSDMVASC09,CJHYWJLWX14,POLMGLFNBY14,HMHCK14,M09,L09a,ZCFK10,S10,SXK10,
L10b,RCKVK12}, electronics \cite{SLZEK11} and acoustics \cite{hamid}. In fact optics and electronics, where
${\cal PT}$-symmetric set-ups can be realized by judiciously balancing gain and loss regions of a system, have provided an excellent playground for experimentally testing many theoretical ideas \cite{RMGCSK10,FXFLOACS13,GSDMVASC09,CJHYWJLWX14,POLMGLFNBY14,HMHCK14,
SLZEK11}. Among these theoretical predictions \cite{L09a}, and subsequent experimental realizations \cite{GSDMVASC09}, was a new type of Bloch Oscillations which were unstable. They either amplified or attenuated since the propagating constants in the associated ${\cal PT}$-symmetric lattices became immediately complex (the system entered the broken ${\cal PT}$-symmetric phase) once a transverse index gradient was introduced.

\begin{figure}[h]
\centerline{\includegraphics[width=8cm]{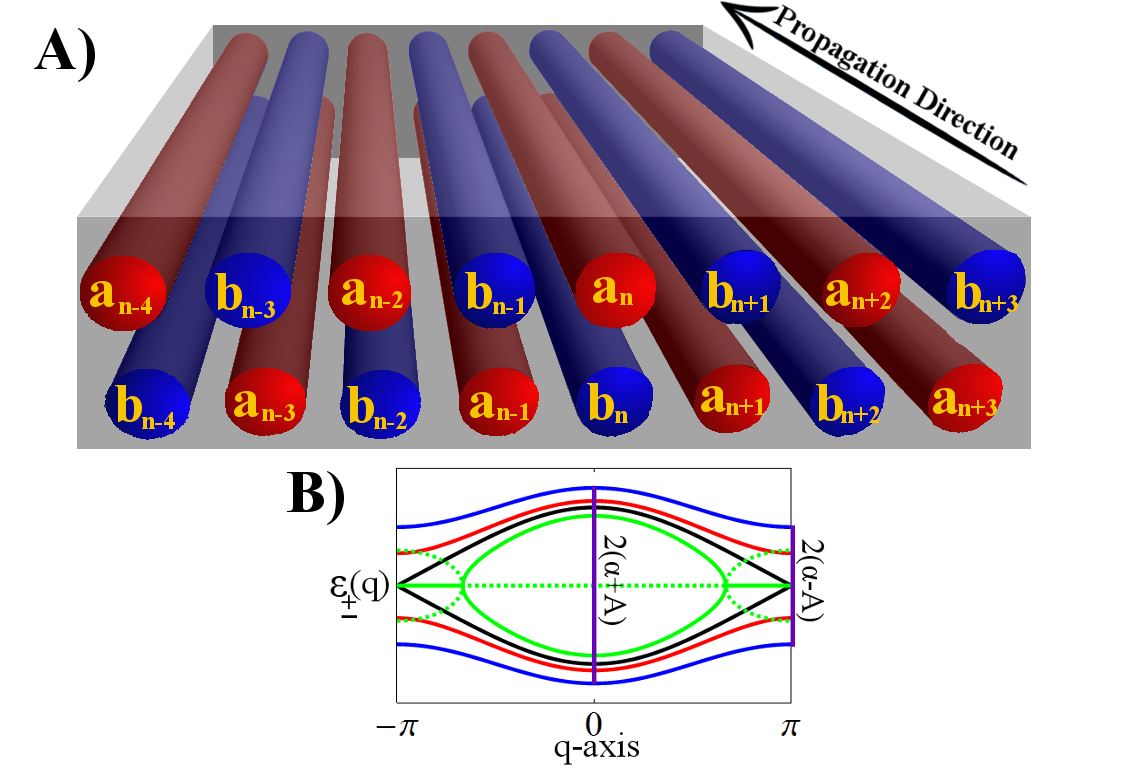}}
\caption{A) The photonic lattice with local ${\cal PT}$-symmetry. B) The associated dispersion relation for $\gamma = 0$ (blue),
$0\leq\gamma\leq \gamma^{\cal{PT}}$ (red), $\gamma =  \gamma^{\cal{PT}}$(black), and the solid/dashed green lines are the real/
imaginary part for $ \gamma> \gamma^{\cal{PT}}$.}
\label{iso}
\end{figure}

In this Letter we introduce a class of photonic lattices,  whose building blocks are ${\cal PT}$-symmetric dimers with a transverse index gradient $\omega$ (see Fig. \ref{iso}a). These (quasi-one-dimensional) lattices respect a {\it local} ${\cal P}_d{\cal T}$-symmetry associated with each individual dimer. Despite the lack of global ${\cal PT}$-symmetry they still have parameter domains for which their eigenvalues are real i.e. they are in the exact ${\cal PT}$-symmetric phase. In this domain they support a new class of {\it stable} ${\cal PT}$-symmetric
Bloch-Zener oscillations which, allow for periodic wavepacket self-imaging whenever the choice of the $\omega-\gamma$ parameters impose a synchronous behavior between the Zener tunneling and the period of Bloch-Oscillations. These synchronous Bloch-Zener oscillations experience a cascade of splittings and giant beam recombinations which are further enhanced in the presence of localized defects.

{\it Theoretical Model--} We consider the photonic lattice of Fig. \ref{iso}a. An experimental implementation of the index gradient for such a
set-up has been realized in Ref. \cite{DSHPNTL09,TPLMSBP06}. Each waveguide supports only one propagating mode, while light is transferred
between waveguides via evanescent tunneling. The connectivity of the array is such that each amplifying (dissipative) waveguide of a dimer
is coupled, with a coupling constant $\frac{A}{2}$, to both of the adjacent dimers' dissipating (amplifying) waveguide. In addition we assume
an intra-dimer coupling $\alpha$. We will assume that $\alpha > A$. The diffraction dynamics of the evolving electric field amplitude
$\Psi_{n}(z) = (a_{n}(z), b_{n}(z))^{T}$ of the $n^{th}$ dimer along the propagation direction $z$, in the paraxial description, satisfies the
following Schr{\"o}dinger-like equation
\begin{equation}
\begin{split}
\label{DIF}
i \frac{da_{n}}{dz}+(n \omega - i \gamma) a_{n} + \alpha b_{n} + \frac{A}{2}(b_{n-1} + b_{n+1})=0\\
i \frac{d b_{n}}{dz}+ (n \omega + i \gamma) b_{n} + \alpha a_{n}+ \frac{A}{2}(a_{n-1} + a_{n+1})=0
\end{split}
\end{equation}
where $a_n (b_n)$ is the field amplitude at the gain (loss) site of the dimer. Although the system described by Eq. (\ref{DIF}) does not respect
a global ${\cal P T}$-symmetry (due to the index gradient), nevertheless there is a {\it local} ${\cal P}_d {\cal T}$ symmetry that it is satisfied
by each individual dimer.

{\it Spectral Analysis--}
It is instructive to start by studying the dispersion relation of the system in the absence of the transverse index gradient i.e. $\omega =0$.
Using the Fourier transformation $a_n(z) =\frac{1}{\sqrt{2 \pi}} \int_{-\pi}^{\pi}{\tilde a}_q(z) e^{i q n} dq$ (similarly for $b_{n}(z)$) Eq.
(\ref{DIF}) takes the form:
\begin{equation}
\label{dynfourier}
i\frac{d}{dz}\left(
\begin{array}{c}
{\tilde a}_q(z)\\
{\tilde b}_q(z)
\end{array}
\right)
= \left(
\begin{array}{cc}
 i\gamma & -v_q\\
 -v_q & -i\gamma
\end{array}
\right)
 \left(
\begin{array}{c}
{\tilde a}_q(z) \\
{\tilde b}_q(z)
\end{array}
\right)
\end{equation}
where $v_q=\alpha + A \cos(q)$. The dispersion relation ${\cal E}^{\sigma}(q)$ (longitudinal propagation constants) is obtained by calculating
the eigenvalues of the $2\times2$ matrix in Eq. (\ref{dynfourier}):
\begin{equation}
\label{dispersion}
{\cal E}_{\sigma}(q)= \sigma \sqrt{(\alpha + A \cos[q])^{2} - \gamma^{2}}
\end{equation}
where $q \in (-\pi,\pi]$ and $\sigma= \pm$ indicates the upper/lower band. For $\gamma =0$ the minimal spacing between the two bands
$\delta = 2 (\alpha - A)$ occurs at $q=\pm \pi$. As $\gamma$ increases the minimal band separation shrinks until the edges touch at $\gamma
\rightarrow \gamma^{\cal{PT}} = \alpha -A$ where an EP degeneracy occurs (see Fig. \ref{iso}b). For $\gamma >\gamma_{\cal{PT}}$ we enter
the broken ${\cal PT}$-symmetric phase and the eigenvalues appear in complex conjugate pairs. Below we will focus our analysis on the parameter
domain for which the spectrum is real (exact ${\cal PT}$-symmetric phase). In this domain, the eigenvectors $\left|\sigma\right\rangle$ of
the $2\times 2$ matrix of Eq. (\ref{dynfourier}) take the form
 \begin{equation}
\left|\sigma\right\rangle =\frac{\sqrt{-\sigma}}{\sqrt{2\cos\theta}}\begin{bmatrix}e^{-i\sigma\theta/2}\\
-\sigma e^{i\sigma\theta/2}
\end{bmatrix};\, \theta=\arcsin\left(-\frac{\gamma}{v_q}\right)
\end{equation}
and they are also eigenvectors of the ${\cal PT}$-operator \cite{BB98}.

When $\omega \neq 0$ the two bands are replaced by two interleaving Wannier-Stark ladders ${\cal E}_n^{\pm}={\cal E}_0^{\pm} +
n\omega$ where $n=0,\pm1,\cdots$. The offsets ${\cal E}_0^{\pm}$ determine the relative energy distance between the two ladders and
can be evaluated numerically from a direct diagonalization of the effective Hamiltonian $H$ that describes the paraxial propagation of our
system Eq. (\ref{DIF}). Moreover, in contrast to the $\omega=0$ case, the system possess multiple exceptional points.

Let us look at the case $A = 0$. In this case the longitudinal propagation constants ${\cal E}_{n}^{\pm}$ are organized in
doublets associated with the $n^{th}$ isolated dimer:
\begin{equation}
\label{LCL}
{\cal E}_{n}^{\pm}= {\cal E}_0^{\pm} + n \omega;\quad {\cal E}_0^{\pm}=\pm \sqrt{\alpha^{2} - \gamma^{2}}
\end{equation}
For $\omega> 2 \alpha$ the spectrum is non-degenerate for any value of $\gamma\neq \alpha$ (for $\gamma=\alpha$ we have multiple EP
degeneracies). However for $\omega = 2 \alpha$ we have a degeneracy at $\gamma = 0$, where ${\cal E}_n^{\pm}={\cal E}_{n\pm 1}^{\mp}$.
Furthermore for $\omega = \alpha$ another (simple) degeneracy develops at $\gamma = 0$ where now ${\cal E}_n^{\pm}={\cal E}_{n\pm 2}^{\mp}$.
At the same time the previous degeneracy at $\gamma=0$ for $\omega = 2 \alpha$, "evolves" towards $\gamma = \alpha \frac{\sqrt{3}}{2}$.
It is straightforward to show that for $\omega_m=\frac{2 \alpha}{m}$, where $m = 1,2,3,...$, degeneracies with more remote dimers occur at $\gamma=0$ while the previous ones evolve towards larger values of $\gamma$. The index $m$, defining the number of degeneracies
for $A=0$, will be used later on in order to delineate the $\omega-A$ parameter space of our system, Eq. (\ref{DIF}), into domains of
broken ${\cal PT}$-symmetry (i.e. number of instability regions) occurring as $\gamma$ increases.

\begin{figure}[h]
\centerline{\includegraphics[width=9cm]{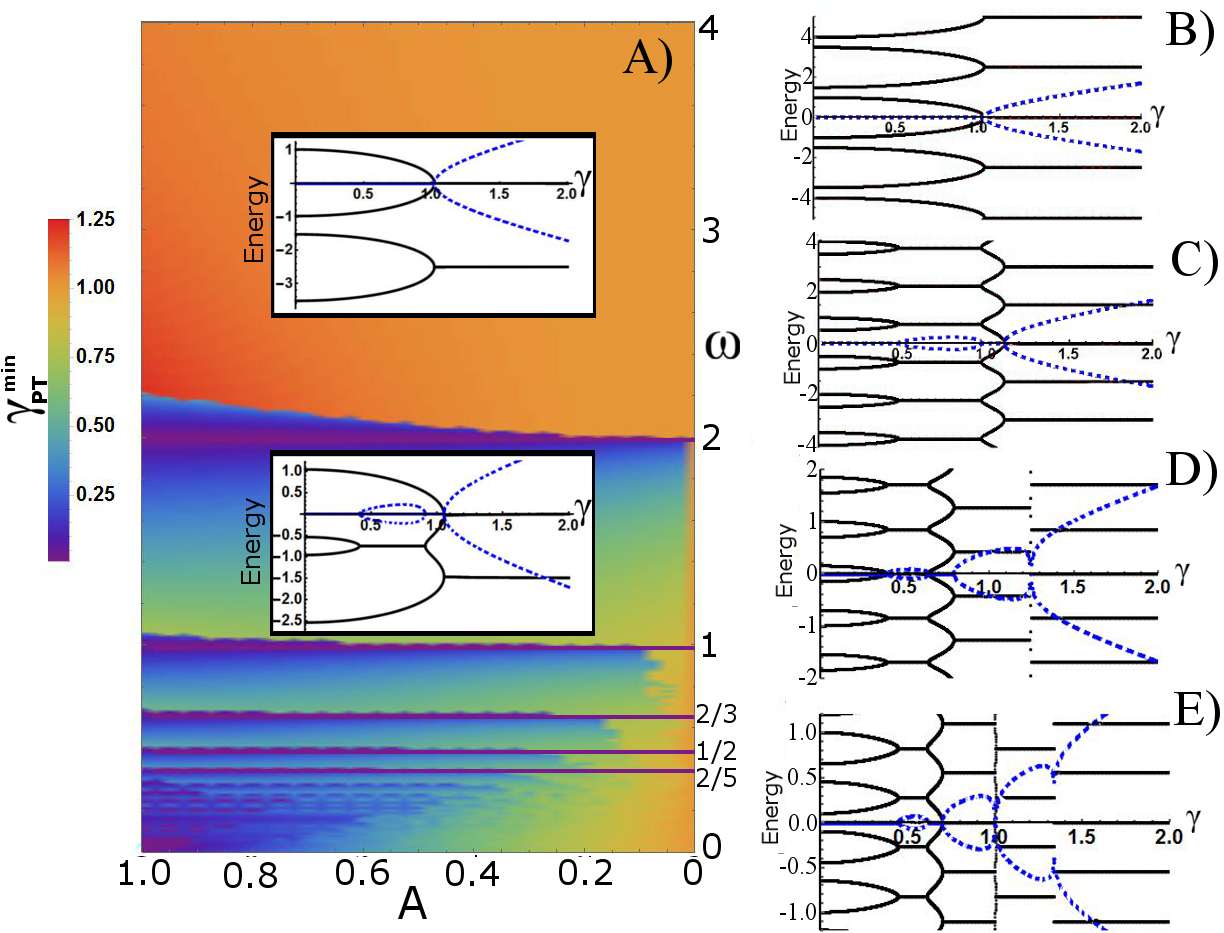}}
\caption{A) Numerical results for the first EP of the system for $\alpha =1$, versus $\omega$ and $A$. B-E) show numerical results for the
real(black) and imaginary (blue) spectra in domains 1, 2, 3, and 4: for $A=0.5$ and $\alpha =1$. In B) $\omega = 2.5$, while in C) $\omega
=1.5$. In D) and E) $\omega = 0.85$ and $\omega = 0.55$ respectively. The sub figures in A) are plots of Eq. (\ref{LCL},\ref{EV}) respectively
for the same parameters as B) and C). The purple lines have been added in A) to illustrate the instability bubbles for small $\gamma$ at the
domain borders}
\label{densityfigure}
\end{figure}

In Fig. \ref{densityfigure}A, we present a density plot for $\gamma_{\cal PT}^{\rm min}$, associated with the first EP, versus $\omega$ and $A$.
The  purple horizontal lines indicate the $\omega_m$-values discussed previously. For each such domain, we plot in Figs. \ref{densityfigure}B-\ref{densityfigure}E, a typical spectral behavior (for fixed $A,\omega$) of the eigenenergies of the system Eq. (\ref{DIF}) versus $\gamma$. We see that the
number of instability regions is described by the index $m$.

The domain $m=1$, associated with one instability region (see Fig. \ref{densityfigure}B), can be understood within the framework of a single dimer,
see Eq. (\ref{LCL}). The latter is also plotted at the inset in Fig. \ref{densityfigure}A. Domain $m=2$, can be analyzed
using two coupled dimers subjected to a gradient $\omega$:
\begin{align*}
H= -& \begin{bmatrix} \bar{n}\omega- i\gamma & \alpha & 0 & \frac{A}{2}\\
\alpha &\bar{n}\omega+ i\gamma & \frac{A}{2} & 0\\
0 & \frac{A}{2} & (\bar{n}-1)\omega - i\gamma & \alpha\\
\frac{A}{2} & 0 & \alpha &(\bar{n}-1)\omega+ i\gamma
\end{bmatrix}
\end{align*}
Direct diagonalization of the above Hamiltonian gives:
\begin{equation}
\label{EV}
{\cal E}_{n}= {\cal E}_0^{\pm}+n\omega;\quad {\cal E}_0^{\pm}=\pm \sqrt{X \pm Y}+{1\over 2}\omega
\end{equation}
where $X \equiv (\frac{A}{2})^2 + \alpha^{2} - \gamma^{2} +(\frac{\omega}{2})^{2}$, and $Y \equiv \sqrt{A^{2} \alpha^{2} + (\alpha^{2}
- \gamma^{2})\omega^{2}}$. Eq. (\ref{EV}) is plotted in the inset of Fig. \ref{densityfigure}a and it describes qualitatively the features (i.e.
two instability domains) shown in Fig. {\ref{densityfigure}c associated with the system Eq. (\ref{DIF}). Other domains $m=3,4,\cdots$ can
be explained by analyzing a system of three, four, etc. coupled dimers. Below we will concentrate only in the parameter space for which
the system is in the exact ${\cal PT}$-symmetric phase (stable domains). We point out again that this feature is absent in Ref. \cite{L09a} where
the system is always in the broken ${\cal PT}$-symmetric phase.


{\it Dynamics}-- To study the dynamics, we have numerically simulated the propagation of a broad Gaussian beam for different values of
$\gamma\leq \gamma_{\cal PT}$. We have assumed a normal incident, so that at the input plane $z=0$ the beam has excited mainly the
first band in a spectral interval around $q_0\approx 0$. We first consider the case of $\gamma=0$ where the
band-gap $\delta=2(\alpha-A)$ is large enough to allow us to neglect Zener tunneling (ZT). According to the acceleration theorem, the
transverse propagation constant $q$ increases up to $\pm \pi$ where the wavelength satisfies the Bragg condition associated with the
underlying periodic potential. The wave is then Bragg reflected at propagation distance $z=\pi/\omega$ and travels in the opposite
transverse direction toward lower index sites where it experiences a total internal reflection. The process repeats itself leading to a periodic motion
which can be considered the optical analogue of Bloch Oscillations. The oscillation period can be easily estimated using the above considerations and it is
$z_B=2\pi/\omega$. The above qualitative picture is nicely reproduced in Fig. \ref{dynamics}A for $\gamma=0$ and $\omega=0.231$.

As $\gamma$ increases the band-gap $\delta$ becomes smaller and ZT between the two bands at their edges $q=\pm \pi$ cannot be
neglected any more. The associated spreading scenario is depicted in Figs. \ref{dynamics}B,C  for $\omega=0.231$ and two different values
of the gain/loss parameter $\gamma=0.405$ and $0.443$. In this case the beam will experience a ZT at distances $z_Z^{(n)}= (2n+1)\pi/
\omega$, where $n=0,1,\cdots$. Let us discuss in more detail the first ZT event at $z_Z^{(0)}=\pi/\omega$. For distances $z<z_Z^{(0)}$
the beam is mainly trapped in the lower band and propagates along the direction of the local gradient $\partial{\cal E}^{-}/\partial q$. At 
$z_Z^{(0)}$, due to the tunneling, the beam splits into two beams one characterized by the lower band and the other by the upper band. 
While the beam associated with the lower band reverses direction via Bragg reflection, the beam associated with the upper band follows a 
parallel trajectory with $\partial{\cal E}^{+}/\partial q$. These two beams will again change direction due to total internal and Bragg reflections 
respectively. They recombine at the second tunneling point at distance $z_Z^{(1)}=3\pi/\omega$. The recombination process is more 
complicated as now both occupied bands experience coherent interference. We have found that at some distances $z_R=z_Z^{(2)}$ (marked 
by the third green line in Fig. \ref{dynamics}C) these recombinations can lead to a giant power focus (the total power is plotted with red line 
in the z-axis of all upper Figs. \ref{dynamics}). The superposition of ZT with Block Oscillations can, in general, result in an asynchronous 
process which destroys exact revivals of the initial packet. Nevertheless we find that wavepacket self-imaging is achieved for some values of
$\omega-\gamma$. This is the case for example for the parameters used in Fig. \ref{dynamics}C (see distance $z_{\rm SI}$
indicated by orange line) as opposed to the results shown in Fig. \ref{dynamics}B where the self-imaging is not observed.

\begin{figure*}[t]
\centerline{\includegraphics[width=\textwidth]{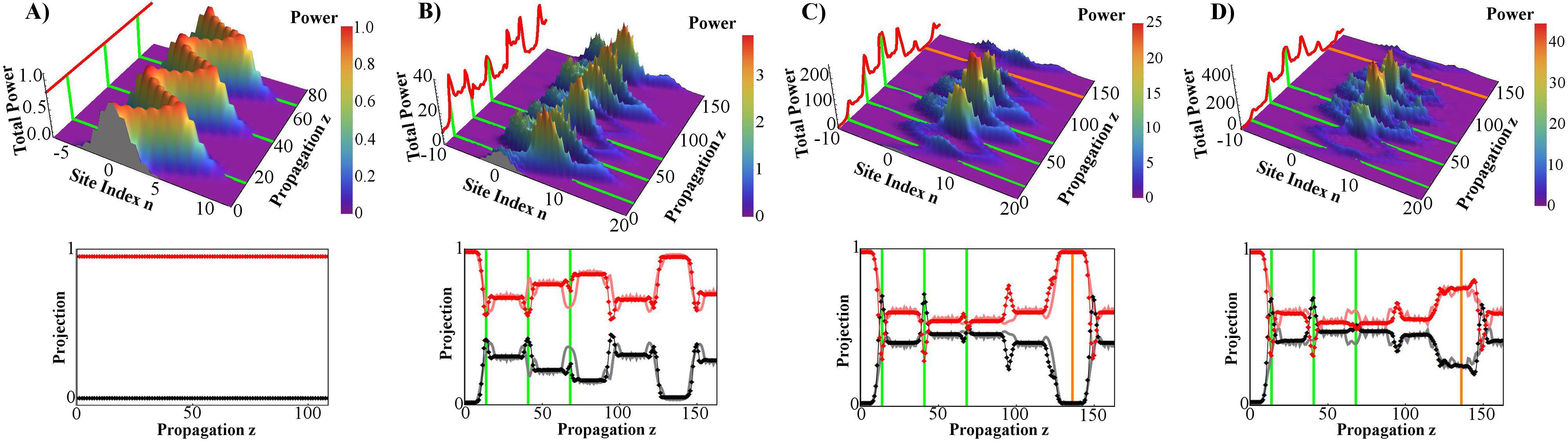}}
\caption{Here we show various dynamical evolutions of the lattice where $\alpha =1$, $\omega=0.231$, and $A = 0.6$, for an initial Gaussian wavefront 
of the form $a_{n}(0)= b_{n}(0)= e^{-n^{2}/10}$. In the upper graphs the x-axis is the site index where the the amplifying site is juxtaposed with the attenuating 
site on the right ($a_{n}(z),b_{n}(z)$) for each index $n$. The red line in the z-axis shows the total power of the lattice while the temperature map color of the 
plot corresponds to the individual site power ($|a_{n}(z)|^{2}, |b_{n}(z)|^{2}$). The green lines mark the first three $z_{Z}$ while the orange lines mark the 
expected self-imaging time $z_{\rm SI}$. In the lower graphs the gray and pink lines correspond to the normalized relative-power on the right and left of the 
recombination point, $n=7$. The black and red dots correspond to the upper and lower band-projections, normalized to one at each instant z. In A) $\gamma 
=0$, in B) $\gamma=0.405$, while in C) and D) $\gamma =0.443$. In D) a defect with strength $\epsilon=0.25$ is included in the dimer with index $n=7$ 
and centered around $z=\frac{\pi}{\omega}$ with a total length of $\Delta z=8$. The presence of the defect disrupts the expected revival at $z_{\rm SI}$
(see the lower subfigure C where $z_{\rm SI}$ is marked with orange line). At the same time it results in a huge power recombination (the power pick is at 
least two times bigger than the one shown in case C) at $z_R$ (indicated by the third green line).}
\label{dynamics}
\end{figure*}


The dynamics is best analyzed in terms of the Floquet-Bloch (FB) eigenvectors of the effective non-Hermitian Hamiltonian
$H$ that describes the paraxial evolution of our system Eq. (\ref{DIF}). Let us indicate, using Dirac's notation, the FB modes associated with the
propagation constant (eigenstate of $H$) ${\cal E}_n^{\sigma}$ as $|{\cal E}_n^{\sigma}\rangle$. They constitute a bi-orthogonal basis and satisfy
the following relations which are dictated by the symmetric nature of $H$
\begin{align}
\left\langle {\cal E}_n^{\sigma *}\right.\left|{\cal E}_{m}^{\sigma'}\right\rangle =\delta_{n,m}\delta_{\sigma,\sigma'};\,\,
\sum_{\sigma=\pm}\sum_{n=-\infty}^{\infty}\left|{\cal E}_{n}^{\sigma}\right\rangle \left\langle {\cal E}_{n}^{\sigma *}\right|={\bf 1},
\label{complete}
\end{align}
where $*$ denotes complex conjugation. Moreover, it is easy to show that the FB modes, in the position space representation
satisfy the following periodicity relation
\begin{align}
\label{6}
\langle \mu,l+k\left|\right.{\cal E}_{n+k}^{\sigma}\rangle = & \langle \mu,l\left|{\cal E}_{n}^{\sigma}\right.\rangle
\end{align}
where $\{\left| \mu, l \rangle \right.\}$ is an orthonormal basis defined by two indexes $(\mu,l)$ with the first index representing the
`gain' $(\mu=1)$ or `loss' $(\mu=2)$ waveguide while the second one denoting the label for the dimer.

Next, we expand the initial preparation $\left|\Psi\left(0\right)\right\rangle$ in the FB basis. The expansion reads $\left|\Psi\left(0\right)
\right\rangle = \sum_{\sigma=\pm}\sum_{n=-\infty}^{\infty}c_{n}^{\sigma}\left|{\cal E}_{n}^{\sigma}\right\rangle$ where $c_{n}^{\sigma}
\equiv \left\langle {\cal E}_n^{\sigma *}\right|\Psi(0)\rangle$. Thus the evolving beam is
\begin{equation}
\left|\Psi\left(z\right)\right\rangle = \sum_{\sigma=\pm}\sum_{n=-\infty}^{\infty}c_{n}^{\sigma} e^{-i{\cal E}_n^{\sigma} z}
\left|{\cal E}_{n}^{\sigma}\right.\rangle .
\label{timeevol}
\end{equation}
We now project the evolving beam Eq. (\ref{timeevol}) to the Wannier-Bloch basis $\left|\sigma,q\rangle
\right.\equiv\left|\sigma\rangle\right.\otimes\left|q\rangle\right.$ where $\left| q \right\rangle=\frac{1}{\sqrt{2\pi}}\sum_{l=-\infty}^\infty
\left|l\right\rangle e^{il q}$ spans the quasi-momentum space:
\begin{eqnarray}
\label{ocup}
\left(\sigma\right.,q\left|\right.\Psi(z)\rangle &=&e^{-i{\cal E}_0^{-}z}\{ C^{-}(\omega z+q)
\left.\left(\sigma\right.,q\right|{\cal E}_{0}^{-}\rangle \\
&+&e^{-i({\cal E}_{0}^{+}-{\cal E}_{0}^{-})z}C^{+}
(\omega z+q)\left(\sigma\right., q\left|{\cal E}_{0}^{+}\right.\rangle\}\nonumber
\end{eqnarray}
where we have used the notation $\left(\sigma,q\left|\right.\right.\equiv \left(\sigma\left|\right.\otimes\langle q\left|\right.\right.$,
and $\left(\sigma\left|\right.=({\cal PT}\right|\sigma\rangle )^T$ \cite{BB98}. Moreover, the coefficients $C^{\sigma}(\omega z+q)
\equiv\sum_{p=-\infty}^{\infty}c_{p}^{\sigma}e^{-ip\left(\omega z+q\right)}$ satisfy the periodicity relation $C^{\sigma}\left(\omega
z+q+2\pi\right)=C^{\sigma}\left(\omega z+q\right)$.

Equations (\ref{timeevol},\ref{ocup}) provide an explanation for the recombination and self-imaging events. They indicate that the
evolving beam is, in general, not periodic as a function of the propagation distance $z$ and it is characterized by two propagation scales:
The first one is the Bloch period, $z_{B} = \frac{2 \pi} {\omega}$, originating from the periodicity of the $C^{\sigma}$ functions. The
second scale $z_{E}= \frac{2 \pi}{{\cal E}_{0}^{+}-{\cal E}_{0}^{-}}$ is associated with the minimal energy spacing in-between the two
Wannier-Stark ladders and arises from the nontrivial relative phase appearing in front of the second term on the rhs of Eq. (\ref{ocup}).
There are $\omega-\gamma$ values for which these two propagation-scales are rationally related to one another i.e.
$z_E/z_B=N/M$. This condition leads to a self-imaging of the initial preparation at propagation distances $z_{SI}=M z_E=N z_B$. For
instance, when $M=1,N=5$ the initial wavepacket is reconstructed at the propagation distance $z_{SI}=\frac{10 \pi}{\omega}$, see the 
orange line in Fig. \ref{dynamics}C . Moreover, a giant power focus (third green line) occurs at the recombination
event which is between two successive self-imaging events.

A deeper insight of the cascade of recombination events can be achieved by evaluating the band contributions of the evolving beam.
Using Eq.~(\ref{ocup}), the band contribution $P^{\sigma}(z)$ can be calculated as 
\begin{equation}
\label{project}
P^{\sigma}(z)\equiv \int_{-\pi}^{\pi} |\left(\sigma\right.,q\left|\right.\Psi(z)\rangle |^2 dq
\end{equation}
The band projections are plotted on the lower row of Fig. \ref{dynamics} as black (upper) and red (lower) points, where we normalize 
$P^{-}(z)+P^{+}(z)=1$. In the same figures the pink and grey lines correspond to the relative power (normalized $P^{R}(z)+P^{L}(z)=1$) 
on the left and right of the recombination point, dimer index $n=7$, as a function of propagation z. This experimental observable 
strongly correlates with the band projections where the pink (left) and grey (right) lines correspond to the lower and upper bands 
respectively. The distances $z=z_Z^{(n)}$ where they demonstrate an oscillatory behavior coincide with the position where Zener 
inter-band transitions of power occurs according to the semi-classical picture of splittings and recombinations discussed earlier 
(see green lines). 

We have also investigated the effect of a localized defect in the creation of these intense recombination points. In general, a defect 
will devalue the maximum of the total power; however, a strategically placed defect at one of the recombination distances $z_Z^{(n)}$ 
can lead to further enhancement of the power peak (see third green line at $z_Z^{(2)}$ in Fig. \ref{dynamics}D and associated power 
pick). We interpret this phenomenon as resulting from quasi-momentum randomization due to the scattering from the defect prior 
to the Bloch-Zener recombination. This leads to a very focused recombinations with all power concentrated in a very narrow lattice domain.

In conclusion, we have investigated stable Bloch-Zener oscillations in a non-Hermitian lattice with local ${\cal PT}$-symmetry. We have found that
an initial beam experiences a cascade of beam splittings and recombinations where the re-concentrated power can exceed the initial value
due to the non-Hermitian nature of the dynamics. At the same time we have found that a judicial selection of the index gradient
$\omega$ and the gain/loss parameter $\gamma$ can result in perfect self-imaging of the initial packet at distances  dictated
by these two parameters. This platform can open up new possibilities for the realization of reconfigurable beam splitters, interferometers
and imaging processing.


\end{document}